\documentclass[11pt]{article}

\usepackage[utf8]{inputenc}
\usepackage[letterpaper, margin=1.25in]{geometry}
\usepackage{amssymb, amsmath}
\usepackage{graphicx}
\usepackage[labelfont=bf]{caption}
\usepackage[colorlinks=true, allcolors=blue]{hyperref}
\usepackage[noabbrev, capitalize, nameinlink]{cleveref}
\usepackage{physics}
\usepackage{enumitem}
\usepackage[export]{adjustbox}
\usepackage[toc, page, header]{appendix}
\usepackage{adjustbox}

\usepackage{natbib}
\hypersetup{citecolor=blue}
\crefdefaultlabelformat{#2\textbf{#1}#3}
\crefname{figure}{\textbf{Figure}}{\textbf{Figures}} 
\crefname{table}{\textbf{Table}}{\textbf{Tables}}
\crefname{appendix}{\textbf{Annexe}}{\textbf{Annexes}}
\hypersetup{citecolor=blue}

\usepackage{titling}
\newcommand{\subtitle}[1]{%
  \posttitle{%
    \par\end{center}
    \begin{center}\large#1\end{center}
    \vskip0.5em}%
}

\title{Socio-Economic Impacts of COVID-19 on Household Consumption and Poverty}
\date{\normalsize \today}

\usepackage{authblk}
\author[1]{Amory Martin\footnote{Corresponding author: Amory Martin, \href{mailto:amorym@stanford.edu}{amorym@stanford.edu}
}}
\author[1]{Maryia Markhvida}
\author[2]{St\'{e}phane Hallegatte}
\author[2]{Brian Walsh}
\affil[1]{\small \textit{Stanford University}}
\affil[2]{\small \textit{Global Facility for Disaster Reduction and Recovery}} 

\providecommand{\keywords}[1]
{
  \noindent \small{	
  \textbf{\textit{Keywords: }} #1}
}

\usepackage[
    type={CC},
    modifier={by-nc-sa},
    version={4.0},
]{doclicense}

\begin{document}

\maketitle

\begin{abstract}

\noindent The COVID-19 pandemic has caused a massive economic shock across the world due to business interruptions and shutdowns from social-distancing measures. To evaluate the socio-economic impact of COVID-19 on individuals, a micro-economic model is developed to estimate the direct impact of distancing on household income, savings, consumption, and poverty. The model assumes two periods: a crisis period during which some individuals experience a drop in income and can use their precautionary savings to maintain consumption; and a recovery period, when households save to replenish their depleted savings to pre-crisis level. The San Francisco Bay Area is used as a case study, and the impacts of a lockdown are quantified, accounting for the effects of unemployment insurance (UI) and the CARES Act federal stimulus. Assuming a shelter-in-place period of three months, the poverty rate would temporarily increase from 17.1\% to 25.9\% in the Bay Area in the absence of social protection, and the lowest income earners would suffer the most in relative terms. If fully implemented, the combination of UI and CARES could keep the increase in poverty close to zero, and reduce the average recovery time, for individuals who suffer an income loss, from 11.8 to 6.7 months. However, the severity of the economic impact is spatially heterogeneous, and certain communities are more affected than the average and could take more than a year to recover. Overall, this model is a first step in quantifying the household-level impacts of COVID-19 at a regional scale. This study can be extended to explore the impact of indirect macroeconomic effects, the role of uncertainty in households' decision-making and the potential effect of simultaneous exogenous shocks (e.g., natural disasters). 

\end{abstract}

\keywords{COVID-19, socio-economic impact, household consumption, poverty rate}

\vspace{0.3in}
\doclicenseThis
\normalsize

\newpage 

\section{Introduction}

COVID-19 has led to severe and acute losses in many economies around the world due to illness and  and government-mandated social distancing orders. The impact and duration of the economic crisis on individual households, resulting from the pandemic, is difficult to predict as many uncertainties surround the crisis duration, i.e. length of ''stay-at-home" orders, as well as impacted industries and the post-crisis consumption and recovery.

There is a plethora of ongoing research studies on estimating the economic impact of COVID-19, in both emerging and developed countries. Due to widespread business closures, especially in lower income populations, national economies are expected to contract, leading to a dramatic rise in unemployment and poverty rates. A report from the World Bank estimated that 11 million people could fall into poverty across East Asia and the Pacific \citep{WorldBank2020EastCovid-19}. Analyzing the effect of the pandemic on poor communities across four continents, \cite{Buheji2020TheReview} estimates that 49 million individuals will be driven into extreme poverty in 2020 (living on less than \$1.90 per day). 

The U.S. economy, where gross domestic product (GDP) fell by 4.8\% in the first quarter, is projected to fall into recession in 2020, with a contraction of 5.0\% in a likely scenario \citep{McKibbin2020TheScenarios, Fernandes2020EconomicEconomy}. The European Commission estimates that the euro area economy would decline by 7.25\% in 2020, with all countries expected to fall into a recession \citep{EuropeanCommission2020European2020}. Developing countries in South-East Asia are also vulnerable to the global economic disruption of the pandemic due to decrease in trade, foreign investment and tourism. According to the International Monetary Fund (IMF), the ASEAN-5, which consists of Indonesia, Malaysia, Philippines, Thailand, and Vietnam is predicted to decline by 0.6\% in 2020 \citep{InternationalMonetaryFund2020WorldLockdown}. Reduction in remittances from high-income countries to low- and middle-income countries is likely to have a significant impacts in many countries, such as Nepal or the Philippines, where remittances represent a large share of many households' income.

In the six week span of March 15 to April 25, a record 30.2 Americans have filed for unemployment benefits as first-time claimants, according to the U.S. Department of Labor. The unemployment rate in the U.S. hit a staggering 14.7\% officially in April from statistics released by the U.S. Bureau of Labor Statistics and some predictions estimate even higher unemployment rates, above 20\%, \citep{Bick2020RealOutbreak}. 

According to the Pew Research Center, the highest risks of layoffs are in the accommodations, retail trade, transportation services and arts entertainment and recreation services sector\citep{Kochhar2020YoungJobs}. Additionally, among the sectors that lost the most jobs in March are the leisure and hospitality and health and educational services \citep{Burns2020HowDemographics}. Using a variable vector autoregression model based on data from recent disasters, \cite{Ludvigson2020Covid19Disasters} estimates a cumulative loss of 24 million jobs in the U.S. over the course of 10 months, largely due to a 17\% loss in service sector employment. Only 37\% of jobs in the U.S. can be performed at home, and many lower-income countries have a lower share of jobs that can be performed remotely \citep{Dingel2020HowHome}. Consumer discretionary spending is in free fall as non-essential businesses are closed and individuals are saving more. Analyzing data from a personal finance website, \cite{Baker2020HowPandemic} found that consumer spending in the United States is highly dependent on the severity of the disease's outbreak in the state and the strength of the local government's response. 


Although ongoing research is assessing the economic ramifications of COVID-19, most of these studies are focused on the macroeconomic and financial impact of the coronavirus. Impact on national economies are then translated into socio-economic impact on individuals, including consumption and poverty rates (top-down approach). The goal of this study is to analyze the socio-economic impacts of the COVID-19 containment at the household level (bottom-up approach). While this approach is not expected to replace macro-level analyses that can better capture the interaction across sectors and countries or the effect of macroeconomic aggregated, it can complement it by providing much finer estimates of the distributional impacts. It can also better account for households' coping capacity, the role of people's savings, and the higher resilience of multi-job households.

To understand the impacts of the loss of revenue on the lower income level populations, a household well-being formulation is adopted following the work of \cite{Hallegatte2016Unbreakable:Disasters}. The original household well-being model was developed for the disaster impact of an earthquake, and applied to the Bay Area in California \citep{Markhvida2020QuantificationLosses}. In addition, the household model has also been applied to estimate household-level resilience to natural disasters in Fiji, the Philippines, and Sri Lanka \citep{Walsh2018ClimateResilient, Walsh2019SocioeconomicAssessment, Walsh2020MeasuringLosses}. Here the economic shock of COVID-19 is represented by loss of income, in certain industry sectors, during a pre-defined crisis period. The impact of the coronavirus on household consumption, savings and recovery time is analyzed, as well as changes poverty rates and geospatial inequality distributions, with different assumptions regarding the social protection system. Since California has been affected early and high-frequency data on the situation of households are available in the U.S., the Bay Area is a good case study to develop and validate the model, which we then plan to apply in other countries and regions.

\section{Case Study}

To evaluate the socio-economic impact of COVID-19 on individuals, a micro-economic model is
built to estimate the household consumption and well-being. The model is used to quantify the
effects of mandatory “shelter-in-place” orders, as well as the effectiveness of social benefits.
The San Francisco Bay Area is used as a case study, where impact of the lockdown and the
recently passed CARES Act federal stimulus package are evaluated in conjunction with state
unemployment insurance (UI) benefits. The model and approach can be easily transferred to other countries 
and regions, even though differences in the availability of data (e.g., regarding financial savings)
may make it necessary to make further approximations in other contexts. 

\subsection{Scope and Data}

\begin{figure}[h!]
    \centering
    \includegraphics[width=0.9\linewidth]{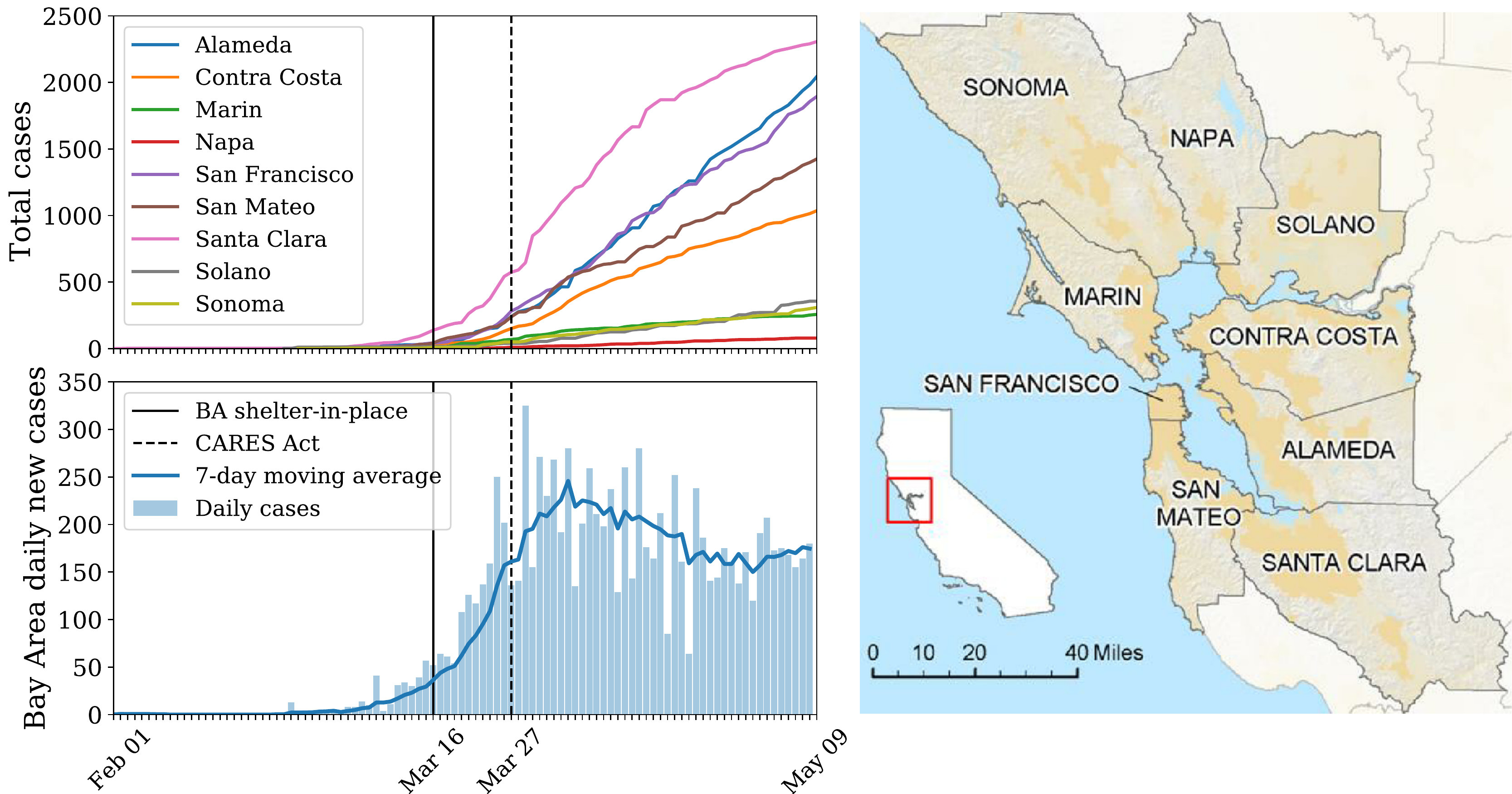}
    \caption{Bay Area COVID-19 crisis timeline: total reported cases by county (top left) and daily reported cases from \textit{The New York Times} publicly available Coronavirus (COVID-19) Data in the United States \citep{TheNewYorkTimes2020AnU.S.} (bottom left) and Bay Area counties \citep{Rissman2008TheDatabase} (right)}
    \label{fig:BayAreaCases}
\end{figure}

The Bay Area enacted a shelter-in-place order on March 16, 2020 in six counties: San Francisco, Santa Clara, San Mateo, Marin County, Alemeda County, Contra Costa. Soon after, a mandatory stay-at-home order across California was issued on March 19, 2020. This major business perturbation has led to a sharp rise in unemployment and severe economic repercussions \citep{Schwartz2020NowherePandemic}. On March 27 2020, the Coronavirus Aid, Relief and Economic Security (CARES) Act was signed into U.S. law, which among other stimulus measures, extends unemployment benefits and gives single payouts to individuals. The number of reported total cases per county across Bay Area, as well as the daily cases are shown in Figure \ref{fig:BayAreaCases}.

For the case study of the Bay Area, the socio-economic impacts of COVID-19 of the following 9 counties are modeled, in alphabetical order, (1) Alameda, (2) Contra Costa, (3) Marin, (4) Napa, (5) San Francisco, (6) San Mateo, (7) Santa Clara, (8) Solona and (9) Sonoma County (Figure \ref{fig:BayAreaCases}). The total population of the Bay Area is greater than 7,302,000 inhabitants. The household data is sourced from census tracts information using the \texttt{SimplyAnalytics} platform, which details income, investments, savings, employed and other relevant data for the year 2016 \citep{SimplyAnalytics2016CensusData}.

\subsection{Income Shock}

An income loss schedule is implemented according to industry sectors to model the shock of the COVID-19 crisis on households from economic impacts related to illness, layoffs and loss of activity due to social distancing orders. The income drop per sector is modeled according to Table \ref{tab:IndustryShock} using the 15 aggregated industry sectors from the U.S. Bureau of Economic Analysis (BEA). The hardest hit industries are assumed to be construction, retail trade, transportation, arts and entertainment \citep{Leatherby2020HowMoney}. Affected individuals are assumed to have a 100\% loss of labor income during the shelter-in-place order, which in this study is referred to as the crisis period. 

\begin{table}[h!]
    \centering
    \caption{Percent of affected individuals for each aggregated industry sector of the U.S. Bureau of Economic Analysis (BEA)}
    \begin{adjustbox}{width=0.85\textwidth}
    \small
    \begin{tabular}{c c p{4in} c} \hline
    \textbf{No.} & \textbf{Sector} & \textbf{Description} & \textbf{Affected}$^\dagger$ \\ \hline
    1 & AGR & Agriculture, forestry, fishing, and hunting & 0\% \\
    2 & MIN &  Mining & 0\% \\
    3 & UTI &  Utilities & 0\% \\
    4 & CON &  Construction & 50\% \\
    5 & MAN & Manufacturing & 10\% \\
    6 & WHO & Wholesale trade & 10\%  \\
    7 & RET & Retail trade &  50\% \\
    8 & TRA & Transportation and warehousing & 50\% \\
    9 & INF & Information & 10\% \\
    10 & FIN & Finance, insurance, real estate, rental and leasing & 10\% \\
    11 & PRO & Professional and business services & 10\% \\
    12 & EDU & Educational services, health care and social assistance & 10\% \\
    13 & ART & Arts, entertainment, recreation, accommodation and food services & 80\% \\
    14 & OTH & Other services, except government & 80\% \\
    15 & GOV & Government & 0\% \\ \hline 
    \multicolumn{4}{l}{$\dagger$: Percent of individuals in sector affected, income drop is assumed to be 100\% for affected pop.} \\ \hline \hline
    \end{tabular}
    \end{adjustbox}
    \label{tab:IndustryShock}
\end{table}

\subsection{Policy Impacts}

To investigate the impacts of policies on per capita consumption and well-being, the following three case studies A, B and C are explicitly considered:
\begin{itemize}[leftmargin=*]
    \item \textbf{Case A: Base.} This is the initial base case, where no unemployment insurance nor stimulus benefit package are considered. In this case, households will smooth their consumption during the crisis by using their savings. 
    \item \textbf{Case B: UI.} This is the Unemployment Insurance (UI) case, where the regular California UI benefits are considered. In this case, individuals who lose their job can receive between \$40 and \$450 per week for a maximum duration of 26 weeks (6 months). According to California Law, an individual can claim UI benefits from a minimum gross income of \$900/quarter (\$300/month). The maximum UI benefit is capped for an individual earning \$11,676/quarter (\$3,892/month) or more.
    \item \textbf{Case C: CARES.} This case considers regular California UI benefits in addition to the new CARES Act stimulus package, signed into law on March 27, 2020. Although the Coronavirus Aid, Recovery and Economic Security (CARES) Act contains many programs, two specific aspects are explicitly modeled here: 
    \begin{enumerate}
        \item Unemployment Insurance Extension
        \begin{enumerate}
            \item \textit{Pandemic Emergency Unemployment Compensation.} Eligible individuals, who exhaust their regular California state UI benefits, can receive up to an additional 13 weeks (3 months) of UI benefits at the original rate, for a total UI state benefits of 39 weeks (9 months).
            \item \textit{Pandemic Unemployment Compensation.} Eligible individuals will benefit from an additional \$600/week flat rate of UI benefit until July 31 on top of the UI and Pandemic Emergency Unemployment Compensation. This unemployment assistance is a flat rate and does not depend on prior income. 
        \end{enumerate}
        \item Stimulus Checks \\
            The U.S. government is issuing direct payments to most Americans, up to \$1,200 through the U.S. Internal Revenue Service (IRS). The stimulus checks are based on annual gross income of the 2018 tax filing year. Individuals earning \$75,000 or less per year will receive \$1,200. Individuals earning more than \$75,000 will receive checks reduced by \$50 for every additional \$100 above that threshold. Individuals having a gross yearly income over \$99,000 will not receive a stimulus check. Couples filing jointly and additional benefits for children dependents under the age of 16 (\$500 per dependent child) are not explicitly modeled due to lack of data.
    \end{enumerate}
\end{itemize}

As for all analysis of post-disaster support and social protection impact assessments, it is critical to consider the practical implementation of the measures, and to include in the analysis unavoidable exclusion errors. In this model, the exclusion error is estimated as 40\%, which is based on 2019 unemployment insurance rates in California according to the Employment Development Department (EDD) (Appendix \ref{sec:CaliforniaLabor}). To investigate the importance of implementation of response measures, various assumptions on this parameter are explored. Excluded individuals are assumed not to receive state UI nor CARES due to ineligibility, claim processing errors, exhaustion of UI benefits, or other reasons. Undocumented immigrants, which represent 9.0\% of the total labor force in California according to the Pew Research Center \citep{Passel2016SizeRecession}, are explicitly modeled and are also expected to receive no state nor federal benefits. The unemployment benefits, from state and federal levels, as well as the stimulus checks are given to individuals with a random time delay representing the delay from the onset of the crisis to the benefits arriving in individual bank accounts. The time delays for each are modeled randomly and independently assuming a lognormal distribution with mean six weeks and standard deviation of three weeks.

\subsection{Poverty Levels}

To investigate the impact of COVID-19 on lower income populations, two poverty levels are considered: (1) poverty, corresponding to the Low Income Level (LIL) defined by the Department of Housing and Urban Development (HUD) and (2) deep poverty, which is defined as half the income of LIL. According to HUD, LIL is defined for household gross annual income less than \$25,844, thus, deep poverty is defined for gross annual income less than \$12,922. These poverty measures closely match the California Poverty Measure (CPM), developed by the Public Policy Institute of California \citep{Bohn2019PovertyCalifornia}. From the census tract data, the poverty rate of the Bay Area is 17.1\%, and the deep poverty rate is 1.68\%.  \\

\subsection{Modeling}

Using the census tract data, a household-level economic model is built, divided into two periods: (1) crisis period, which simulates the duration of the shelter-in-place order and subsequent loss of income and (2) the recovery period. During the crisis period, affected individuals will suffer an income loss (Table \ref{tab:IndustryShock}) and use precautionary savings to replenish consumption. The income shock is assumed to begin on the first day of the crisis and last until full economic reopening. During recovery, income is assumed to be fully replenished to pre-crisis levels and savings are replenished using a marginal savings rate of 10\%. A household recovery time is defined as the time it will take to fully replenish its savings to pre-crisis levels. Unemployment insurance, from state and federal levels, replenish consumption, while the CARES single paycheck is assumed to replenish savings directly. The methodology, modeling and optimization are presented in full detail in Appendix \ref{sec:Method}.

\section{Household Consumption and Saving Losses}

In this section, the crisis is assumed to last three months, $T_C = 3$, representing a time period starting from the Bay Area shelter-in-place order on March 16, 2020 to a full reopening on June 16, 2020. At the time of writing this report, the California mandated shelter-in-place is to be maintained until at least end of May. However, the order could remain in effect for longer since the daily new cases in the Bay Area is still close to peak levels (Figure \ref{fig:BayAreaCases}). 

\begin{figure}[h!]
    \centering
    \includegraphics[width=\linewidth]{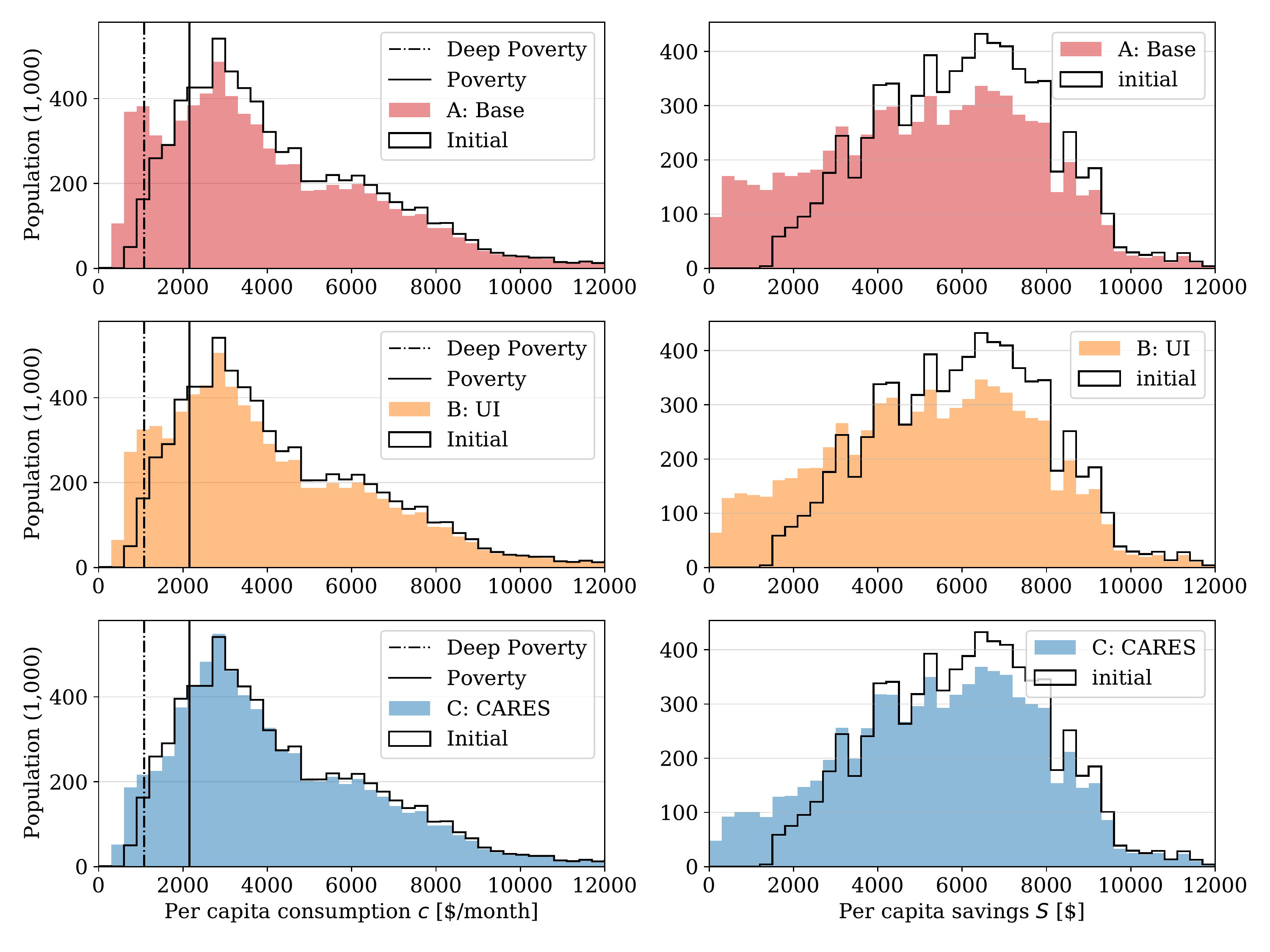}
    \caption{Histograms of per capita consumption and savings, comparing initial pre-crisis and during the crisis period for (a) case A: base, (b) case B: unemployment insurance and (c) case C: unemployment insurance and CARES Act stimulus. The income thresholds for poverty and deep poverty are plotted for comparisons.}
    \label{fig:ConsumptionSaving}
\end{figure}

The histograms of initial versus crisis per capita consumption [\$/month] and savings [\$] for each case (A: base, B: UI and C: CARES) along with the poverty income levels are shown in Figure \ref{fig:ConsumptionSaving}. The median Bay Area initial per capita consumption is \$3,989 per month. Considering consumption during the crisis period and no unemployment insurance (UI) benefit nor federal assistance, the consumption drops for most individuals, with 643,000 people falling below the poverty level  (per capita consumption lower than o \$2,154 per month). The California state unemployment insurance (UI) benefits help maintain consumption during the crisis, with CARES (Case C) having a very strong impact on consumption levels during the crisis.

Considering both unemployment insurance and CARES benefits, the standard deviation of crisis consumption across individuals is reduced. This is represented by the lower tail distribution of Figure \ref{fig:ConsumptionSaving}(c). Indeed, lower income individuals gain more from CARES Act then their job revenue pre-crisis, since the \$600/week unemployment benefit is an average and not based on per capita income.

The median per capita savings in the Bay Area, before the crisis is \$6,092, which represents 7.0 weeks of pre-crisis consumption. With no benefits (case A), most individuals deplete their savings to smooth consumption, with some individuals fully using the precautionary savings. For case B, considering UI benefits, the decrease in savings is reduced thanks to California benefits. Finally, the residual savings are much higher with both UI and CARES (case C), since the state and federal benefits are used as alternative cash flows instead of using savings to replenish consumption. 

\section{Recovery Time}

\begin{figure}[h!]
    \centering
    \includegraphics[width=0.5\linewidth]{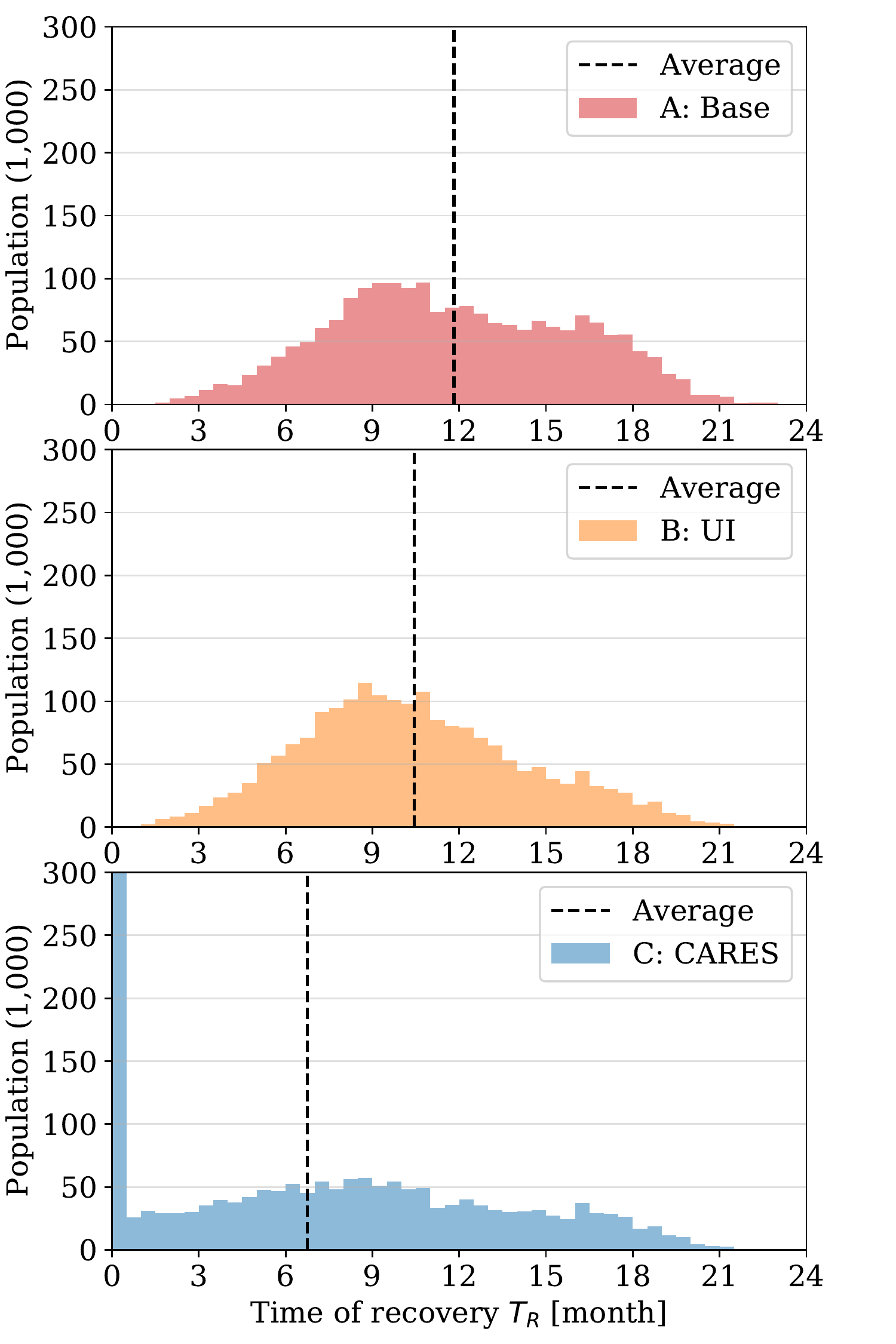}
    \caption{Histograms of recovery time for affected populations considering (a) case A: base, (b) case B: unemployment insurance and (c) case C: unemployment insurance and CARES Act stimulus.}
    \label{fig:RecoveryTime}
\end{figure}

The histograms of recovery time in months for the affected population (i.e., those who have an income loss due to the COVID-19 crisis) in the Bay Area is shown in Figure \ref{fig:RecoveryTime} along with the average values. The recovery time is defined as the time needed to replenish savings, which were depleted during the crisis. The crisis period is assumed to last three months. 

 The median and per-quartile recovery times for affected individuals is shown in Table \ref{tab:AvgRecoveryTime}. The average recovery time for case A is almost a year (11.8 months), which illustrates the severity of the economic crisis due to COVID-19. The average recovery time in case B is reduced by 3 months as a result of UI state benefits to an average of 10.4 months, since individuals replenish their savings faster. The CARES Act (case C) further reduces the average recovery time to 6.7 months, but also reduces its depth. Most individuals have a recovery time less than half a month in this situation, even though the situation is very heterogeneous.  Although the median recovery time dramatically decreases with the addition of CARES, the third quartile (Q3) does not drop as much. Indeed, 25\% of affected individuals will take more than 11 months to fully recover from the crisis.
 
 \begin{table}[h!]
    \centering
    \caption{Median and quartile recovery times in months for affected individuals.}
    \begin{tabular}{l c c c c} \hline
        \textbf{Cases} & \textbf{A: base} & \textbf{B: UI} & \textbf{C: CARES} \\ \hline 
        Q1         &8.7     &7.7     &0.0 \\
        median    &11.5    &10.1     &6.3 \\
        Q3        &15.1    &12.9    &11.2 \\ \hline \hline
    \end{tabular}
    \label{tab:AvgRecoveryTime}
\end{table}
 
\begin{figure}[h!]
    \centering
    \includegraphics[width=0.7\linewidth]{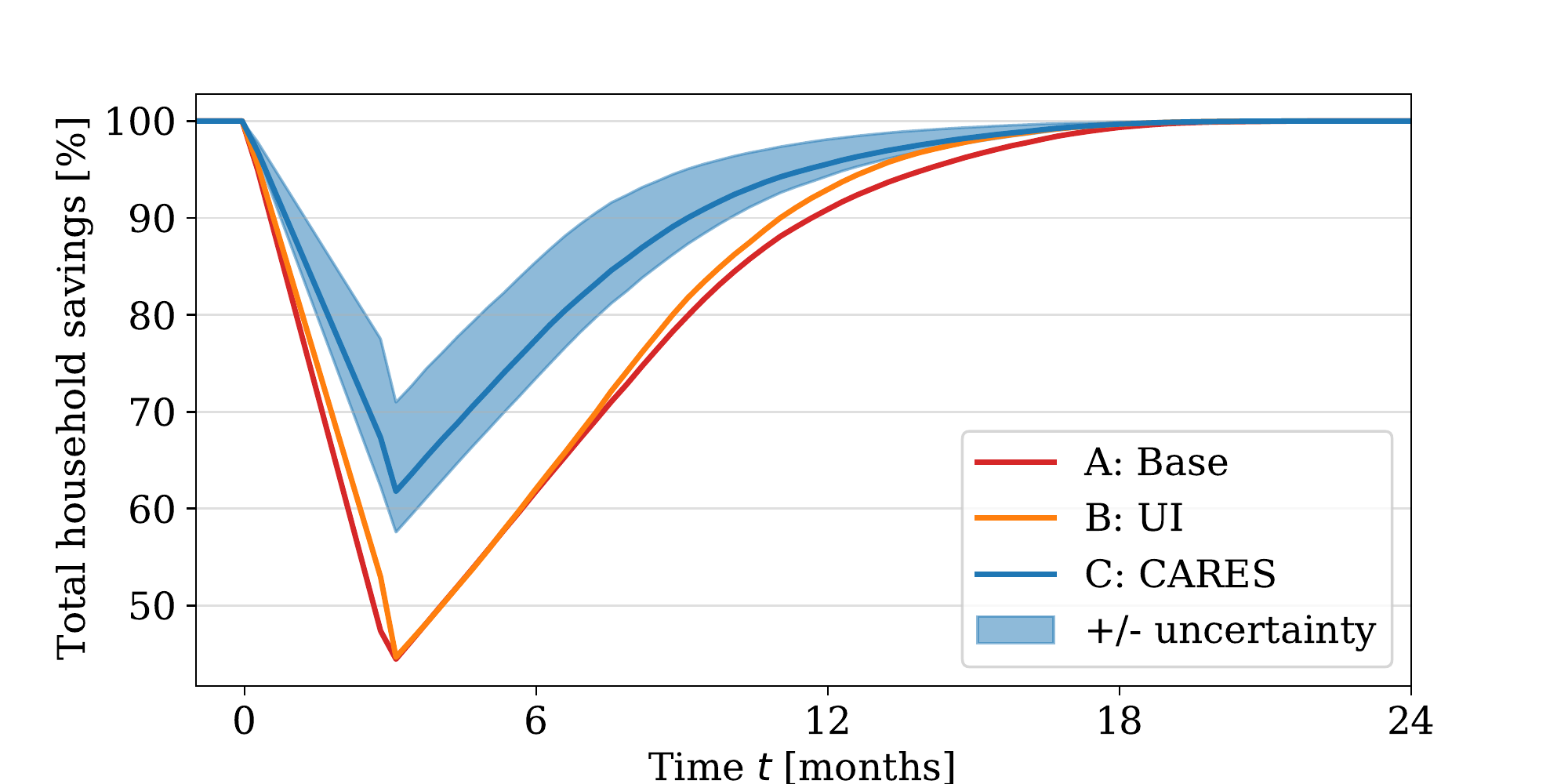}
    \caption{Recovery curve for the Bay Area using total household savings as a function of time for cases A (base), B (UI) and C (CARES). For case C, a confidence interval is shown based on uncertainty in exclusion rate (55\% to 10\%).}
    \label{fig:RecoveryCurve}
\end{figure}

 Altogether, the full recovery takes more than 12 months in all cases, as illustrated by Figure \ref{fig:RecoveryCurve}. The figure shows total household savings as a percentage of pre-crisis level. The recovery remain long because, even with the most optimistic assumptions, there are some individuals who takes a long time to fully recover. 

Moreover, it is important to note that this model assumes that individuals regain full employment and income as soon as the crisis is over, and it does not account for the longer-term macroeconomic repercussions of the pandemic crisis. In reality, the drop in demand while households rebuild their asset (and firms rebuild their balance sheet) and the uncertainty in the timeline of the COVID pandemics is likely to maintain income depressed for a long time period. Introducing this macroeconomic feedback will be a priority for future work. 

\section{Poverty and Policy Impact}

\begin{figure}[h!]
    \centering
    \includegraphics[width=1.0\textwidth, center]{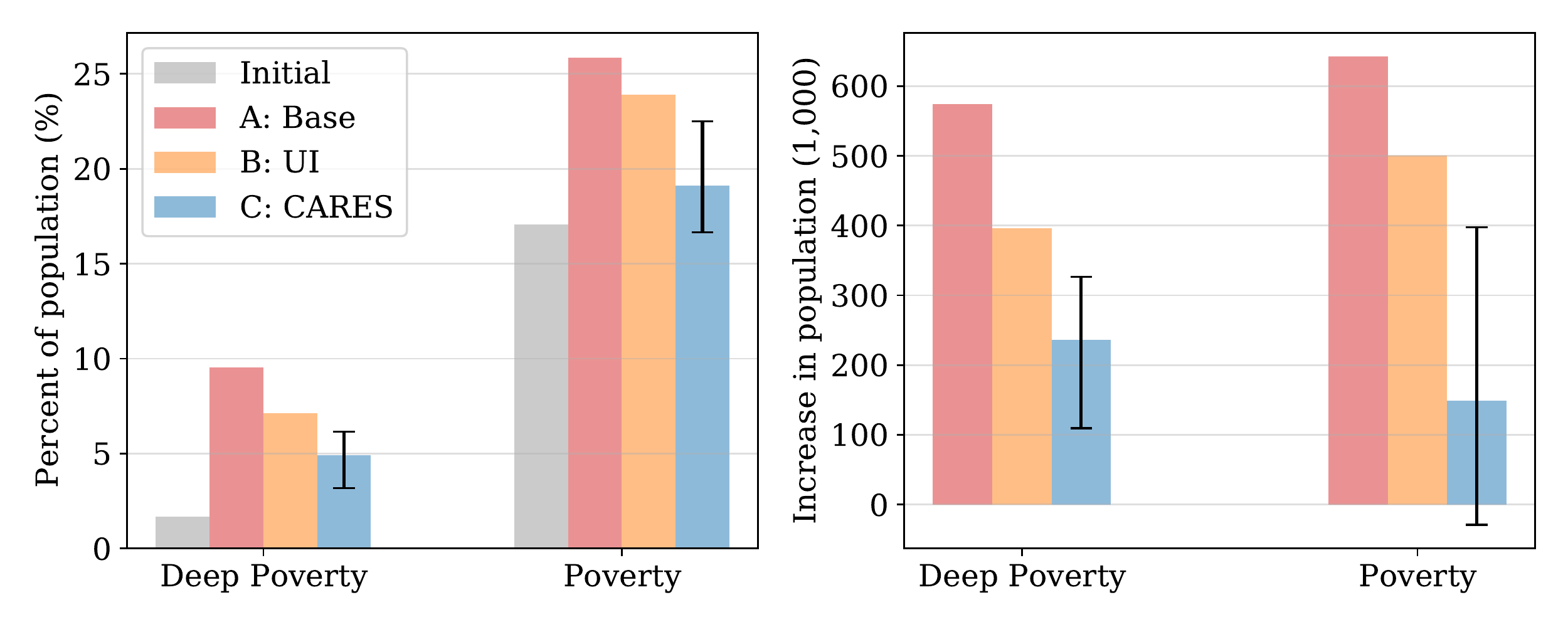}
    \caption{Bay Area poverty and deep poverty rates, as well as increase in poverty populations considering initial, case A, case B and case C. For case C with CARES, uncertainty is given based on exclusion rate with a median of 40\% and upper and lower bounds of 55\% and 10\% respectively.}
    \label{fig:IncomeLevels}
\end{figure}

In this section, the impact of different policies on lower income individuals are evaluated. The following lower income levels are used to analyze the policies' impact: poverty level and deep poverty levels. Figure \ref{fig:IncomeLevels} shows the deep poverty and poverty rates, as well as the increase in number of individuals under those poverty levels. The poverty rates for cases A, B and C are computed based on consumption levels at the end of the crisis, and represent temporary poverty rates, as compared to annual poverty rates as is traditionally reported. For case C, uncertainty in implementation is accounted for using a median exclusion rate of 40\% in addition to worst-case and best-case scenarios of 55\% and 10\%. 

Under case A, the deep poverty and poverty rates temporarily increase dramatically from 1.7\% to 9.5\% and 17.1\% to 25.9\% respectively, from pre-crisis to post-crisis levels. With no social protection, the Bay Area could have an additional 643,000 people in poverty. Cases B, and C reduce the increase in poverty rates by providing income and benefits to impoverished individuals. Under case C, with an assumed 40\% exclusion rate, the deep poverty rate drops from 9.5\% assuming no social benefits, to 4.9\% with CARES. The poverty rate would still increase by 2.0\% points even with the implementation of state and federal assistance. 

\begin{table}[h!]
    \centering
    \caption{Pre and post-crisis deep poverty and poverty rates in the Bay Area.}
    \begin{tabular}{l c c c c} \hline 
        &\textbf{Initial}  &\textbf{A: base}  &\textbf{B: UI}  &\textbf{C: CARES} \\ \hline 
        Deep poverty      &1.7\%     &9.5\%     &7.1\%     &4.9\% [2.5\% - 6.2\%]$^\dagger$ \\
        Poverty          &17.1\%    &25.9\%   &23.9\%    &19.1\% [15.7\% - 20.8\%]$^\dagger$ \\ \hline
        \multicolumn{5}{l}{$\dagger$: confidence interval on exclusion rate [55\% - 10\%]} \\ \hline \hline
    \end{tabular}
    \label{tab:PovertyRate}
\end{table}

\begin{figure}[h!]
    \centering
    \includegraphics[width=\linewidth]{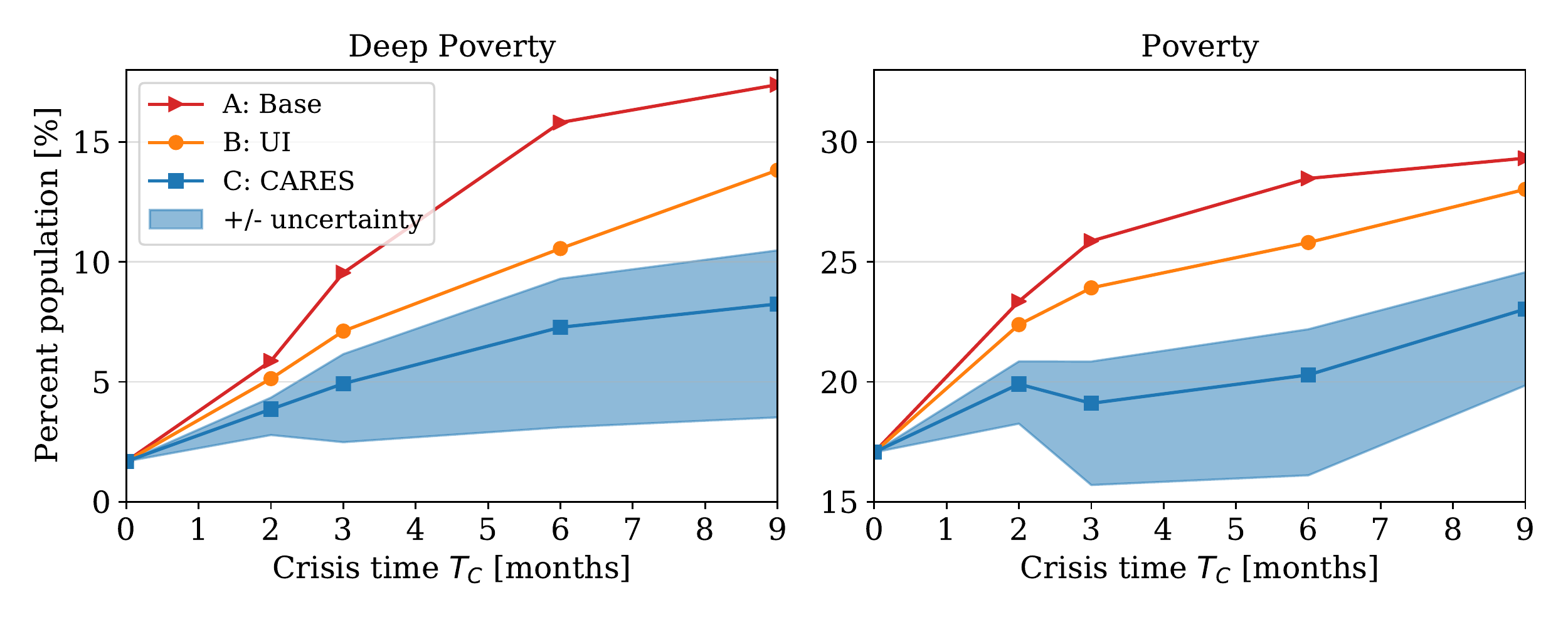}
    \caption{Impact of crisis time on deep poverty and poverty rates in the Bay Area for cases A, B and C. The shaded area for case C represents the uncertainty in the exclusion rate with a likely estimate of 40\% and upper and lower bounds of 55\% and 10\% for worst-case and best-case implementation scenarios respectively.}
    \label{fig:CrisisTimeImpact}
\end{figure}

Given the uncertainty around the duration of the crisis period, the effect of the different policies are evaluated for different crisis periods: $T_C = $ 2, 3, 6 and 9 months. Figure \ref{fig:CrisisTimeImpact} illustrates the percent of the Bay Area population under each lower income level for all three policies evaluated using the four different crisis periods. Cases B and C reduce the number of individuals for each lower income levels. Considering cases B and C, the time of the crisis exacerbates the financial strains on the Bay Area population, reflected by the increase in percentage of the population for each lower income level. 

Most notably, case C, considering both UI and CARES benefits, reduces the number of individuals at the poverty levels, which a minimum at 3 months. This is because the \$600/week extra UI benefits expire at approximately 4 months. Thus, individuals who are laid off and earn less than the UI benefit, will get a bump in their income and consumption. Indeed, in the short-term, some individuals may benefit from staying unemployed for up to 4 months, thanks to the higher income from social benefits. 

\section{Inequality and Geospatial Distribution}

\begin{figure}[h!]
    \centering
    \includegraphics[width=\linewidth]{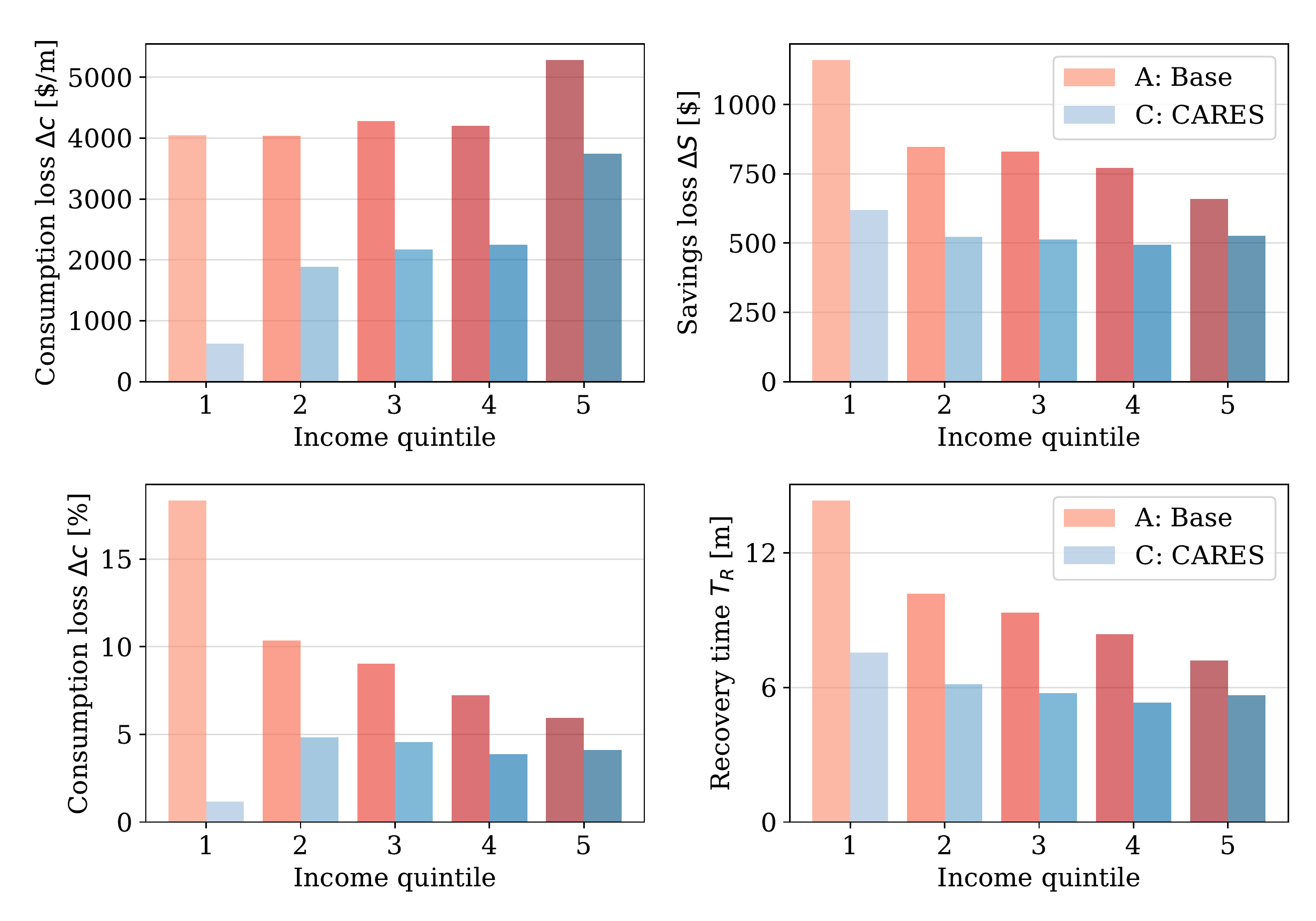}
    \caption{Household average consumption losses, total [\$/month] and relative [\%], saving losses and recovery time by income quintile for cases A (base) and C (CARES) assuming a crisis period of three months.}
    \label{fig:QuintileA}
\end{figure}

The average household consumption losses, both total [\$/month] and relative [\%], saving losses and recovery time (for affected individuals) by pre-crisis income quintile are shown in Figure \ref{fig:QuintileA} for cases A and C as a comparison, assuming a crisis period of three months and a 40\% exclusion rate for unemployment insurance and CARES. For the base case A with no assistance, although the total consumption losses [\$/month] are higher for the higher income individuals, the reverse is true considering relative income losses [\%]. The lowest income quintile has an average relative consumption loss of 18.3\%, compared to only 5.9\% for the highest income earning individuals. Furthermore, the average recovery time for affected individuals is double for the lowest income quintile compared to the highest income quintile, 14.3 compared to 7.2 months. Without social protection, the lowest income population is most impacted by the coronavirus crisis.

On the other hand, with government assistance (case C), the consumption losses are smaller for all income quintiles, with the lowest drop for the lowest income individuals. Most likely, average consumption losses are the smallest for the lowest quintile since benefits from the assistance program can be superior to pre-crisis income in certain cases. In addition, unemployment insurance and the federal stimulus package lead to a more equal distribution of average saving losses and recovery time. 

\begin{figure}[h!]
    \centering
    \includegraphics[width=\linewidth]{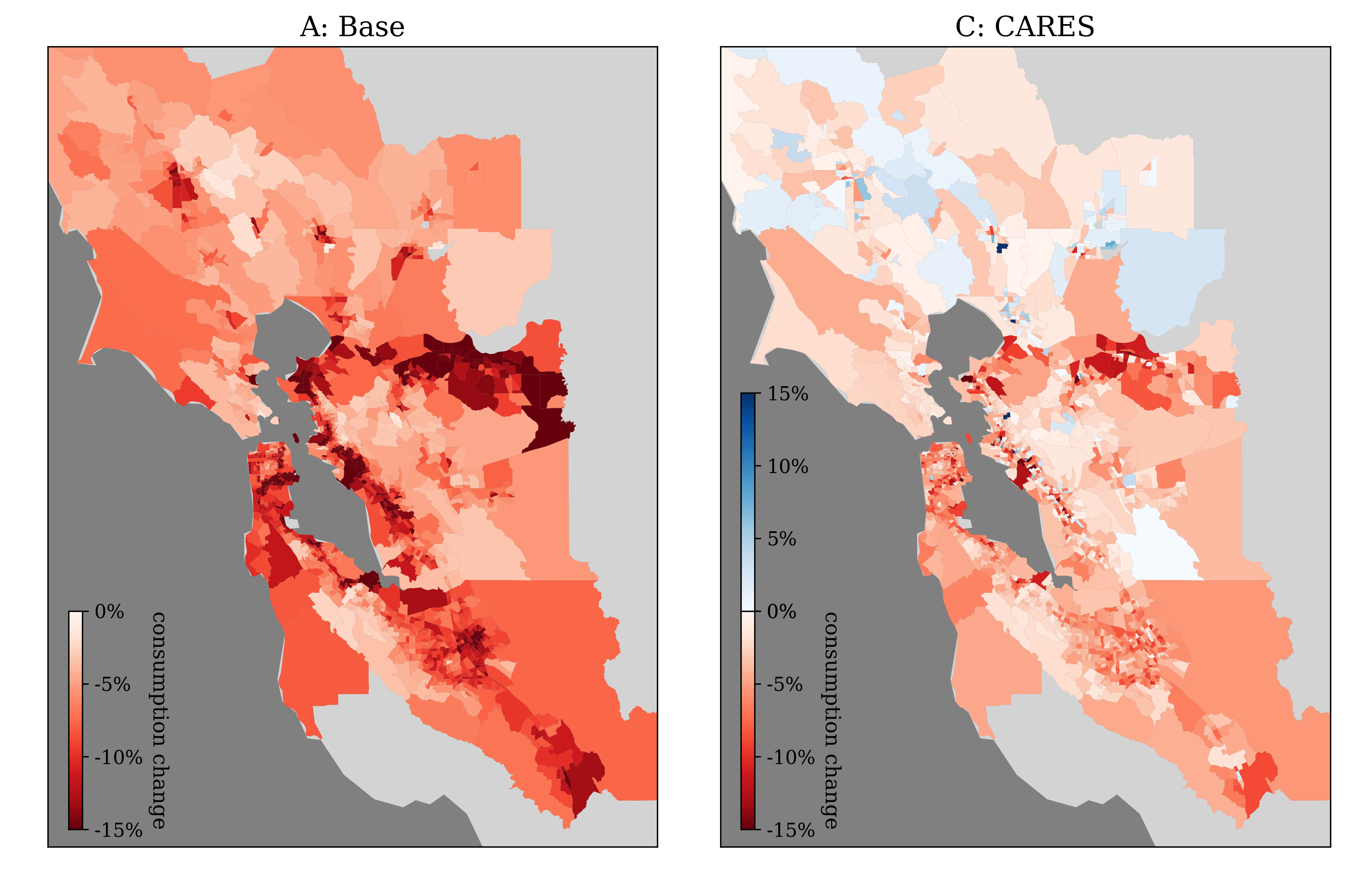}
    \caption{Spatial distribution of average relative consumption change [\%] per capita in the Bay Area for case A: no benefits (left) and case C: CARES (right). Red indicates an average consumption loss (negative values) and blue indicates an average consumption gain (positive values).}
    \label{fig:ConsumptionLossesMap}
\end{figure}

The geospatial distribution of average household consumption losses [\%] in the Bay Area is shown in Figure \ref{fig:ConsumptionLossesMap} for cases A (base) and C (CARES). Overall, the economic impacts of the business interruptions due to the coronavirus are felt throughout all Bay Area counties. The effects of COVID-19 are particularly felt in Alameda and Contra Costa counties, where the average consumption losses exceed 15\% in multiple regions. Cities of South San Francisco, Richmond, San Leandro and Concord are particularly affected. In comparison, the addition of unemployment insurance and individual benefits from the CARES Act lead to a lower average consumption loss across all counties in the Bay Area, assuming a 40\% exclusion rate for case C. Certain regions of the Bay Area even see a moderate increase in consumption, due to the social benefits of CARES. Contra Costa and Alameda counties remain the hardest hit areas with consumption losses greater than 10\% for multiple regions.  

\begin{figure}[h!]
    \centering
    \includegraphics[width=\linewidth]{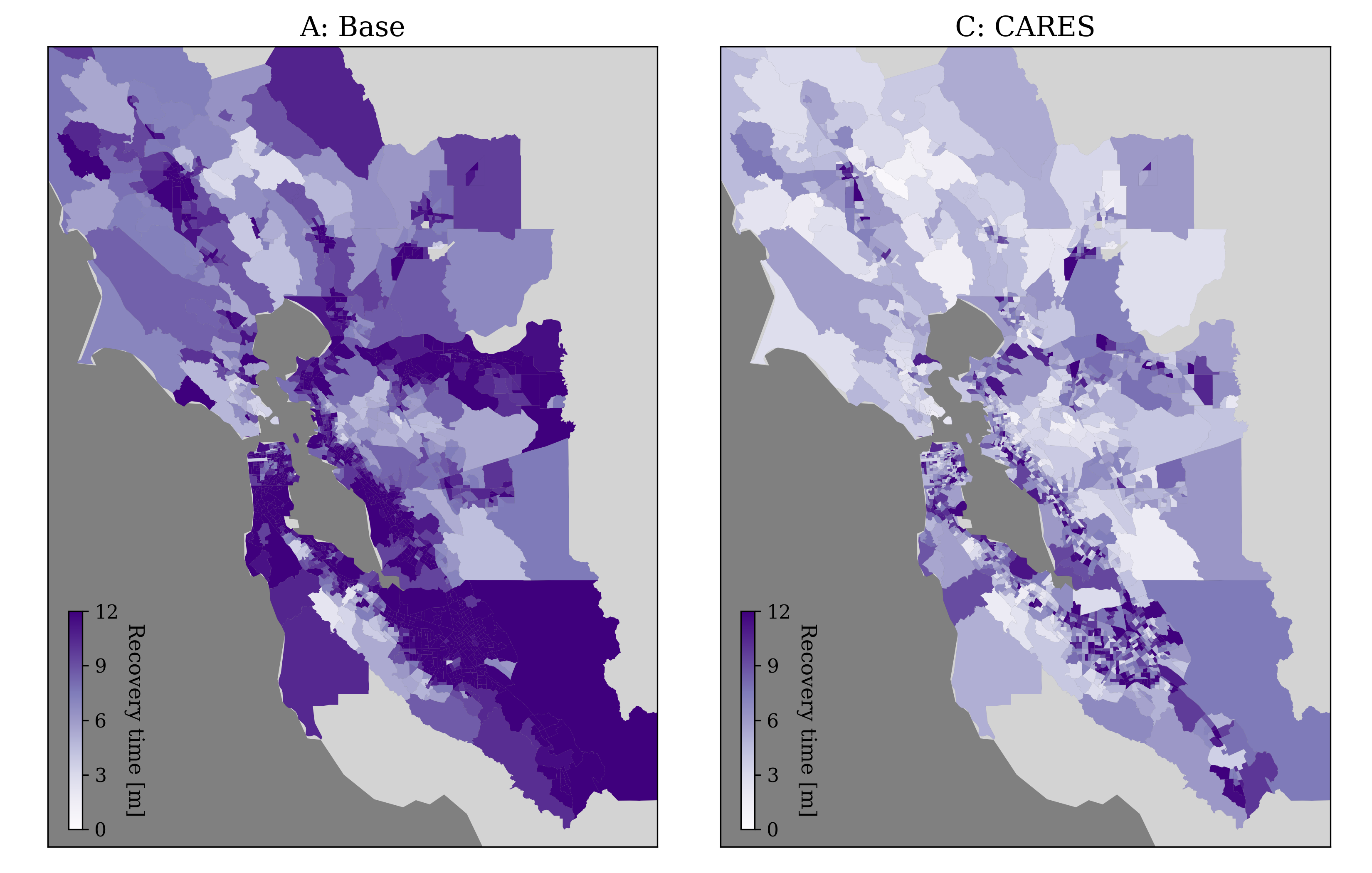}
    \caption{Spatial distribution of average recovery time [months] per capita for affected individuals in the Bay Area for case A: no benefits (left) and case C: CARES (right). Darker purple indicates a longer recovery time.}
    \label{fig:RecoveryTimeMap}
\end{figure}

The geospatial distribution of average recovery time for affected individuals in the Bay Area is illustrated in Figure \ref{fig:RecoveryTimeMap} considering cases A and C and assuming a crisis period of three months. Considering no social protection, vast regions of the Bay Area have an average recovery time over 9 months, even exceeding 12 months in certain cases. CARES and unemployment insurance significantly diminish the recovery time of most regions in all nine counties of the San Francisco Bay Area. However, the average recovery time of affected individuals is very heterogeneous. Densely populated regions in San Jose, San Francisco and East Bay can see average recovery times exceed a year, while some rural regions, such as those near Marin and Napa Couties drop below three months. 

\section{Assumptions, Limitations and Future Work}

There are several limitations to the modeling presented in this study, specifically, (1) the distribution of income loss by industry sector, (2) the exclusion rate of unemployment insurance and CARES and (3) delay in payments for filed claims. The underlying assumptions, as well as the rationale, are discussed in this section. 

Firstly, the average loss of income by industry sector, according to the broad categories of the Bureau of Economic Analysis (BEA), was implemented to reflect the direct impact of the coronavirus and business interruptions, which particularly hit non-essential service sectors. The hardest hit industry sectors were assumed to be arts, entertainment, recreation, accommodation and food services (BEA Sector 13 ART) and other services except government (BEA Sector 14 GOV), quickly followed by construction (BEA Sector 4 CON), retail trade (BEA Sector 7 RET), transportation and warehousing (BEA Sector 8 TRA), see Table \ref{tab:IndustryShock}. This is based on reports from the Bureau of Labor Statistics (BLS), which estimates that the hardest hit sectors are the hospitality and leisure services, especially the accommodation and food sectors \citep{Franck2020HereImpact}. Out of a survey of 10,000 participants, \cite{Coibion2020TheSpending} found that 50\% of respondents had an income loss with an average of more than \$5,000 due to the coronavirus lockdown. 

However, there is now growing evidence of second-round effects, with unemployment growing in sectors that are not directly impacted by the containment, but by supply chain effects (e.g., affected firms reducing demand for other firms) or the aggregate drop in demand. To estimate these effects, a next step is to connect the household model presented here to an Input-Output model, such as the one used in \cite{Guan2020} to look at how COVID-19 affects global supply chains, or a Computable General Equilibrium (CGE) model. 

Income loss for affected individuals is assumed to start at the beginning of the crisis and lasts the duration of the crisis. At the start of the recovery phase, affected individuals regain full employment and the marginal rate of savings is assumed to be 10\% of consumption. This is a conservative estimate, as many employees who were laid off during the crisis will have difficulties re-entering the economy due to the large macroeconomic effects of the coronavirus and the potential for a recession \citep{Avalos2020CoronavirusLockdowns}. In addition, social distancing measures will progressively be relaxed and some service activities, such as food services, bars and performing arts will only be fully operational in the distant future. Furthermore, household consumption during the crisis is assumed to be constant, where households distribute income revenue, state and federal assistance and savings to replenish consumption. In reality, the crisis might exacerbate savings depletion at the onset due to the delay in assistance.

Under the assumed distribution of income loss by sector, and their representative share of the Bay Area, the unemployment rate is estimated at 27.4\% due to coronavirus by the end of the crisis. Certain estimates, such as the calculations from the Federal Reserve Bank of St. Louis, using filed unemployment claims from the Bureau of Labor Statistics (BLS), place the unemployment rate in the U.S. as much as 32.1\% by the end of Q2 \citep{Faria-e-Castro2020Back-of-the-EnvelopeRate}. As of April 23, the Labor Department reported that 26 million people in the United States have filled for unemployment insurance in solely the last five weeks \citep{Cohen2020JoblessCrisis}, 3.4 million of these are in California \citep{Avalos2020CoronavirusLockdowns}. From the Bureau of Labor Statistics (BLS), the California workforce is 19,485,000 as of 2019 with 731,000 unemployed individuals, representing a 3.9\% unemployment rate. Adding the 3.4 million claimants yields a 21.2\% unemployment rate in California, but this only includes the individuals who have filed for unemployment as of April 23, 2020 and the number is likely to grow by the end of the crisis. Here we are assuming that the nine counties Bay Area are approximately representative of the overall California state.

Secondly, the exclusion rate of state and federal assistance (CARES) is assumed to be 40\%. However, there are many uncertainties on this rate, considering the unprecedented scale of the stimulus package. With our median assumption, 40\% of all unemployed individuals will not receive state and/or federal unemployment assistance during the crisis. This is due to issues related to eligibility and to implementation challenges, such as erroneous data or system backlog. According to the Employment Development Department (EDD) of the State of California, the insured unemployment rate (IUR) is 2.19\% (Appendix \ref{sec:CaliforniaLabor}), while the unemployment rate is 3.9\% in 2019. This means that in 2019, before the crisis, the exclusion rate was 43.8\%. Other statistics estimate an exclusion rate as high as 59\% in California \citep{Badger2020StatesUndo.}. Here, it is assumed that the state UI exclusion rate will stay constant during the constant. 

The rate may however increase as a result of number of people losing their job. It could also go down, since the CARES Act introduced the Pandemic Unemployment Assistance (PUA) program to boost the percentage of unemployed individuals receiving benefits, especially gig-workers, freelances, contractors and self-employed individuals. However, as of April 10, the EDD of California has not received guidelines from the federal level on how to implement the program \citep{Castaneda2020AllCoronavirus}. 

Thirdly, the delay in state and federal assistance is assumed to follow a lognormal distribution with mean six weeks from the start of the crisis and standard deviation of three weeks. In reality the delay in stimulus checks could be much higher, individual who file taxes without linking their banking information with the IRS, could receive checks up until September. Due to the unprecedented volume of 26 million claims across the union in just five weeks, the processing of claims and disbursement of federel unemployment insurance (UI), as well as single payment stimulus checks is likely to be severely delayed \citep{Hernandez2020CoronavirusApplications, Wire2020NewlyBenefits}. Widespread website glitches, system crashes and backlog on phone help lines have exacerbated the problem and contributed to lag in UI payments being issued \citep{Badger2020StatesUndo.}. 

There are many potential extensions and applications of this study. Firstly, the long-term macroeconomic impacts of COVID-19 were mainly ignored, but in reality, could lead to income depressed for a longer period of time than the crisis. The income drop could spread to other industry sectors that feel the secondary effect of the loss of mainly the service and hospitality sectors. Additionally, the role of uncertainty in households decision-making could change the rate of savings depletion and the severity of the impact of the crisis. Households do not have perfect information about the duration nor depth of the crisis. Finally, the impact of simultaneous exogenous shocks, such as natural disasters, is of great concern, since lower income populations will have depleted most of their savings and are vulnerable to another shock. For instance, Puerto Rico experienced a 5.4 magnitude earthquake on May 2 during the lockdown with several buildings damaged. Although mild, this scenario highlights the potential for a real crisis due to an added shock.

\section{Conclusion}

In this study, we propose a household-level model to assess the socio-economic impacts of COVID-19 on per capita consumption and savings, and the benefits from government interventions. Assuming an income loss distribution for various sectors, the model can provide estimates of households' consumption losses (a proxy for well-being), depletion of precautionary savings, and recovery time. The San Francisco Bay Area was used as a case study. The main findings of this study are the following.

First, without any social protection, COVID-19 would lead to a massive economic shock to the system. In simulations of a 3-month lockdown, the poverty rate increases from 17.1\% to 25.9\% during the crisis in the Bay Area. Household savings and consumption drop significantly, and the average recovery time for individuals is almost one year. The long recovery time after the crisis will be further exacerbated by a general decrease in demand, people’s change in consumption behavior, and general slowdown of economic activities.

Second, government benefits, both state UI and federal CARES stimulus, decrease the amplitude and duration of the crisis. In likely scenario of a 3-month crisis period, the increase in poverty can be limited to 19\% (from 17.1\% at pre-crisis), and the average time of recovery almost halved to 6.7 months, thanks to the state UI and the federal stimulus package. However, the recovery is spatially heterogeneous, as certain communities will be impacted more than the average and could take over a year to replenish their lost savings.

A near perfect implementation of CARES Act, with 90\% of unemployed individuals receiving benefits, could even lead to a slight temporary decrease in the poverty rate in the Bay Area from 17.1\% to 16.5\%, since the unemployment compensation is higher than pre-crisis income for certain individuals.

Further work will explore the impact of indirect and macro-level impacts, the role of uncertainty in households' decision-making and the consequences in case of multiple waves of social distancing and the possible effect in case of simultaneous exogenous shocks (e.g., natural disasters). Indeed, these results are particularly important when considering the risk of multiple shocks: where the COVID-19 crisis is forcing most households to use their precautionary savings (especially in countries with weak social protection system), the population becomes much more vulnerable to any other shocks, including other natural disasters (e.g., tropical storms, with the hurricane season starting in the Caribbean on June 1st, earthquakes, a 5.5 magnitude earthquake hit Zagreb, Croatia on March 22, 2020 during the lockdown) or the financial and economic secondary impact from the expected recession. 

Beyond this first modeling exercise, the model can be used in other countries or regions, and provide assessment of the potential impact from the ``shelter-in-place mandates", as well as the benefits from different options to provide emergency income support. 


\newpage
\bibliography{references.bib}


\newpage
\appendix
\section{Methodology, Modeling and Optimization}
\label{sec:Method}

\subsection{Income, Consumption and Savings}

The initial pre-disaster income, $i_o$, is: 

\begin{equation}
\begin{aligned}
    i_o &= i_o^L + i_o^{oth} + i_o^h \\
    &= i_o^L + \pi k_o^{oth} + \pi k_o^h
\end{aligned}
\end{equation}
where $i_o^L, i_o^{oth}$ and $i_o^h$ are the initial pre-disaster incomes from labor, investments and housing respectively\footnote{Here, what we refer to as housing income is the imputed rent for homeowners, considered as a capital income.}, $k_o^{oth}$ and $k_o^h$ are capital stocks for investments and housing respectively and $\pi$ is the US average productivity of capital. The total income as a function of time, $i(t)$, is defined as follows:
\begin{equation}
\begin{aligned}
    i(t) &= i_o - \Delta i(t) \\
    &= i_o - \Delta i^L(t) + i^{UI}(t) + i^{CARES}(t)
\end{aligned}
\end{equation}
where $\Delta i^L(t)$ is the labor income loss over time due to the crisis and $i^{UI}(t)$ and $i^{CARES}(t)$ are the unemployment insurance income and CARES Act 2020 stimulus package income respectively. The initial pre-disaster household consumption, $c_o$ is:
\begin{equation}
    c_o = i_o - p_o^{rent} - p_o^{mort}
\end{equation}
where $p_o^{rent}$ and $p_o^{mort}$ are the rent and mortgage payments\footnote{Rent and mortgage payment are removed since the housing income is included in $i_o$. As a simplification, it is assumed that all income that is not  invested in housing is being consumed.}. 

Initially, households have also precautionary savings $S_o$, which they can use to smooth consumption in case of income shock. It is assumed that the containment phase last for a duration $T_C$. After this period, incomes can get back to their pre-crisis level, and there is a recovery period of duration $T_R$ during which households rebuild their precautionary savings. 

As a first exploration, this study assumes that there is no macroeconomic-level impact of the crisis: the only impact is a decline in the income of some households, either because they cannot work remotely or because demand has collapsed in their sector. People who are not directly affected through a drop in revenue or loss of job are assumed that have an unchanged income. These assumptions are acceptable over the short-term, but will be increasingly optimistic as the duration of the containment last. Over the longer-term, one can expect all workers and firms to be affected as the impact of reduced incomes propagate through the economic system. These second-round effects will be explored in a second phase.  

During the crisis and recovery period, households use and then rebuild their precautionary savings and the consumption as a function of time, $c(t)$, is as follows:
\begin{equation}
    c(t) = \begin{cases} 
    c_o - \Delta i(t) + \displaystyle \frac{S_o - S_f}{T_C} & \mbox{if } 0 \le t \le T_C \\[10pt]
    c_o - \displaystyle \frac{S_o - S_f}{T_R} & \mbox{if } T_C < t \le T_C + T_R \\
    \end{cases}
    \label{Eq:Consumption}
\end{equation}
where $S_o$ and $S_f$ and the initial and final savings respectively, $T_C$ and $T_R$ are the duration of crisis and recovery respectively. The adjusted income is the following:

\begin{equation}
    c_{adj}(t) = \max(c(t), c_{min})
\end{equation}
where $c_{min}=1e^{-3}$ represents the survival level of consumption, assuming people always have access to humanitarian assistance (e.g., food banks). 

Finally, the household savings as a function of time, $S(t)$, are:
\begin{equation}
    S(t) = \begin{cases} 
    S_o - t \displaystyle \frac{S_o - S_f}{T_C} & \mbox{if } 0 \le t \le T_C \\[10pt]
    S_f + \displaystyle \frac{t - T_C}{T_R} (S_o - S_f) & \mbox{if } T_C < t \le T_C + T_R \\
    \end{cases}
    \label{Eq:Savings}
\end{equation}
where $t$ is the time, which is initialized at the start of the crisis $t_o = 0$, and other terms are defined previously. The CARES stimulus individual paycheck (up to \$1,200) is added with a time delay directly into savings. The recovery time is based on an exogenous ability to save, assumed constant for all households: 

\begin{equation}
    T_R = \frac{S_f - S_0}{\gamma c_0}
\end{equation}
where $\gamma$ is the saving rate during recovery, until precautionary savings are back to their pre-crisis level (here we will assume $\gamma=0.10$).

\begin{figure}[h!]
    \centering
    \includegraphics[width=0.7\linewidth]{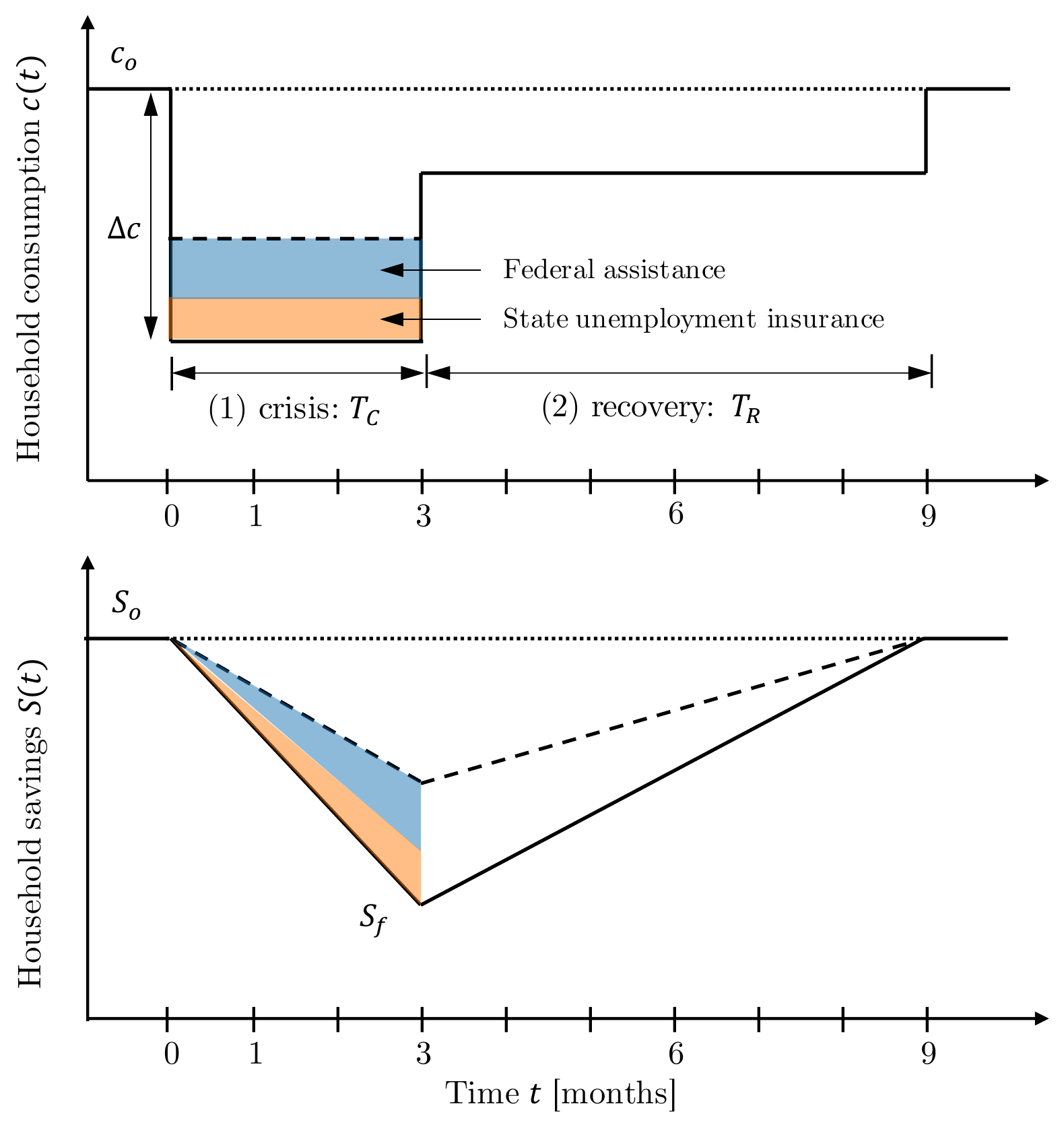}
    \caption{Household consumption and savings model with crisis and recovery periods. Highlighted zones indicate federal assistance and state unemployment insurance}
    \label{fig:Model}
\end{figure}

The model, with household consumption and savings time series, is shown in Figure \ref{fig:Model}.

\subsection{Utility Functions and Household Well-Being}

We assume people derive an utility from consumption, $u(t)$, and from precautionary savings, $v(t)$:\footnote{Utility from precautionary savings can be interpreted either as the value of ``peace of mind'' when people have precautionary savings, or as the net prevent value of the future use of precautionary savings for any shock that can occur in the future.}
\begin{equation}
    u(t) = \frac{1}{1 - \eta} c(t)^{1 - \eta}
\end{equation}
\begin{equation}
    v(t) = \frac{\alpha}{1 - \beta} S(t)^{1 - \beta}
\end{equation}
where $\eta$ is the elasticity of the marginal utility of consumption, $\alpha$ and $\beta$ represent statistically calibrated parameters for utility of precautionary savings (see Appendix). 

The household well-being, $W$, is the sum of the household well-beings during crisis, $W_C$ and recovery phases, $W_R$:

\begin{equation}
\begin{aligned}
    W & = W_C + W_R \\
    &= \int_0^{T_C} e^{-\rho t} \left( u_C(t) + v_C(t) \right) dt + \int_{T_C}^{T_C+T_R} e^{-\rho t} \left( u_R(t) + v_R(t) \right) dt \\
    W & = \int_0^{T_C} e^{-\rho t}  \left( \frac{1}{1 - \eta} c_C(t)^{1 - \eta}
    + \frac{\alpha}{1 - \beta} S_C(t)^{1 - \beta} \right) dt \\
    & \qquad \qquad \qquad  + \int_{T_C}^{T_C+T_R} e^{-\rho t} \left( \frac{1}{1 - \eta} c_R(t)^{1 - \eta} + \frac{\alpha}{1 - \beta} S_R(t)^{1 - \beta} \right) dt \\
\end{aligned}
\end{equation}
where the household consumption $c_C, c_R$ and savings $S_C, S_R$ for the crisis and recovery periods are detailed previously in equations. The household well-being losses, $\Delta W$, are defined as follows: 

\begin{equation}
    \Delta W = W_o - W
\end{equation}
where $W_o$ is the initial well-being defined as follows:
\begin{equation}
\begin{aligned}
    W_o &= \int_0^{T_C + T_R} e^{-\rho t} \left( u_o + v_o \right) dt \\
     &= \int_0^{T_C + T_R} e^{-\rho t} \left( \frac{1}{1 - \eta} c_o^{1 - \eta}
    + \frac{\alpha}{1 - \beta} S_o^{1 - \beta} \right) dt \\
    W_o &= \frac{1}{\rho} \left( 1 - e^{-\rho (T_C + T_R) } \right) \left(  \frac{1}{1 - \eta} c_o^{1 - \eta} + \frac{\alpha}{1 - \beta} S_o^{1 - \beta} \right) dt \\
\end{aligned} 
\end{equation}

The utility at the minimum level of consumption $c_{min}$ provides a lower bound for people's utility. It also provides a higher bound to the well-being impact individuals can experience. Note that this model does not include mortality and morbidity, either due to COVID-19 or to health impacts from containment, such as under- and malnutrition due to insufficient income, mental health implications from isolation, or indirect health consequences from reduced access to health care (especially for people with chronic disease.

\subsection{Optimization Formulation}

In this study, we will assume that households will deplete their savings to smooth consumption over time, in order to maximize their well-being. There are not using all their precautionary savings, because other shock may affect them during or after the COVID-19 crisis, so that the utility derived from remaining precautionary savings have an increasing value as they are used in the current shock. 

One important simplification here is that people are assumed to know in advance the duration of the containment phase. In reality, one challenge for households is to decide how to manage their precautionary savings in the context of a highly uncertain crisis, both in duration and magnitude. The assumption that the duration is known means that the results from the analysis are conservative, underestimating well-being and poverty consequences from containment. 

The one-dimensional unconstrained optimization problem is the following: 
\begin{equation}
    \begin{array}{l  l}
         \text{maximize: } & W(S_f) = W_C(S_f) + W_R(S_f) \\[10pt]
         \text{subject to: } & 0 \le S_f \le S_0 \\ 
    \end{array}
    \label{eq:optimization}
\end{equation}
where $S_f$ is the design variable, the savings at the end of the crisis. The optimization problem is sequentially solved for every census tract in the database representing Bay Area households.

\subsection{Convex Optimization Proof}

In this section, we prove that the optimization problem Eq. \ref{eq:optimization} is convex. Since this is an unconstrained optimization (besides the convex inequality box constraints), to show the problem is convex, we need to prove that $W(S_f)$ is a concave function in $S_f$ (maximization), or $-W(S_f)$ is a convex function (minimization). Thus, we need to show that $-W_C(S_f)$ and $-W_R(S_f)$ are both convex functions. The following Lemmas are necessary for the derivation \citep{Boyd2004ConvexOptimization}:

\paragraph{\textbf{Lemma 1}}
Let $f(x,t):R^2 \rightarrow R$ be a convex function in $x$ for each $t \in [a, b]$, and $w(t) \ge 0$, then the function $\phi$, defined as follows:
\begin{equation}
    \phi(x) = \int_a^b w(t) f(x,t) dt
\end{equation}
is a convex function in $x$.

\paragraph{\textbf{Lemma 2}}
Let $f(x): R \rightarrow R$ be a convex function in $x$ and $a, b \in R$. The function $g$ defined by:

\begin{equation}
    g(x) = f(ax + b)
\end{equation}
is a composition of an affine function, $x \mapsto ax + b$, and $f$, and is convex in $x$. \\

\noindent Thus, since $e^{-\rho t} > 0$ for all $t$, $-W_C$ is convex if and only if $-u_C$ and $-v_C$ are convex functions, similarly for $-W_R$, $-u_R$ and $-v_R$ (Lemma 1). 

Since $1-\eta < 0$, $1-\beta < 0$ and $\alpha>0$, $f(x) = \frac{-1}{1 - \eta} x^{1 - \eta}$ and $g(x) = \frac{-\alpha}{1 - \eta} x^{1 - \beta}$ are convex functions. In addition, since $c(t), S(t)$ are piecewise affine functions in $S_f$, $-u(t), -v(t)$ are the compositions of affine functions and convex functions. By Lemma 2, $-u(t), -v(t)$ are convex functions in $S_f$.

To conclude, $-W_C(S_f)$ and $-W_R(S_f)$ are convex, thus $-W(S_f)$ is a convex function in $S_f$. The optimization problem is convex (QED). \hfill $\blacksquare$

\subsection{Calibration of Utility of Precautionary Savings}

The utility of precautionary savings is of the functional form:
\begin{equation}
    v(t) = \frac{\alpha}{1 - \beta} S(t)^{1 - \beta}
\end{equation}
where $\alpha$ and $\beta$ are parameters to statistically calibrate. Assuming an exponential law relation between consumption and savings: 
\begin{equation}
    S_o = a c_o^b
\end{equation}
where $a, b$ are parameters to calibrate that describe the exponential law. Equilibrium between savings and consumption is assumed initially, before the crisis:

\begin{equation}
    \left. \frac{du}{dc} \right|_{c_o} = \left. \frac{dv}{dS} \right|_{S_o}
\end{equation}
which leads to the following relation $\beta = \eta / b$. 

\noindent Using variational form of the household well-being: 
\begin{equation}
    c_\lambda (t) = c_o + \lambda \delta_o(t)
\end{equation}
where $\lambda$ is an instantaneous increase of consumption and $\delta_o(t)$ is the Dirac delta function. The variational savings is the following:
\begin{equation}
    S_\lambda (t) = S_o + \lambda
\end{equation}

\noindent The variational well-being is now:
\begin{equation}
\begin{aligned}
    W(\lambda, t) &= \int_0^{+\infty} e^{-\rho t} \left( \frac{1}{1 - \eta} c_\lambda(t)^{1 - \eta}
    + \frac{\alpha}{1 - \beta} S_\lambda(t)^{1 - \beta} \right) dt \\
    W(\lambda, t) &= \int_0^{+\infty} e^{-\rho t} \left( \frac{1}{1 - \eta} (c_o + \lambda \delta_0(t))^{1 - \eta} + \frac{\alpha}{1 - \beta} (S_o + \lambda)^{1 - \beta} \right) dt
\end{aligned}
\end{equation}

Taking the derivative in terms of $\lambda$ and setting to zero at equilibrium:
\begin{equation}
    \pdv{W}{\lambda} = \int_0^{+\infty} e^{-\rho t} \left( \delta_0(t)(c_o + \lambda \delta_0(t))^{- \eta} + \alpha (S_o + \lambda)^{-\beta} \right) dt = 0
\end{equation}

\noindent Solving we get the following equation (setting $\lambda=0$ at equilibrium conditions):
\begin{equation}
    \int_0^{+\infty} e^{-\rho t} \left( \delta_0(t) c_o^{-\eta} + \alpha S_o^{-\beta} \right) dt = 0
\end{equation}

\begin{equation}
    c_o^{-\eta} - \frac{\alpha}{\rho} S_o^{-\beta} = 0
\end{equation}

\begin{equation}
    \alpha = \rho \frac{c_o^{-\eta}}{S_o^{-\beta}}
\end{equation}

\begin{figure}[ht]
    \centering
    \includegraphics[width=0.6\linewidth]{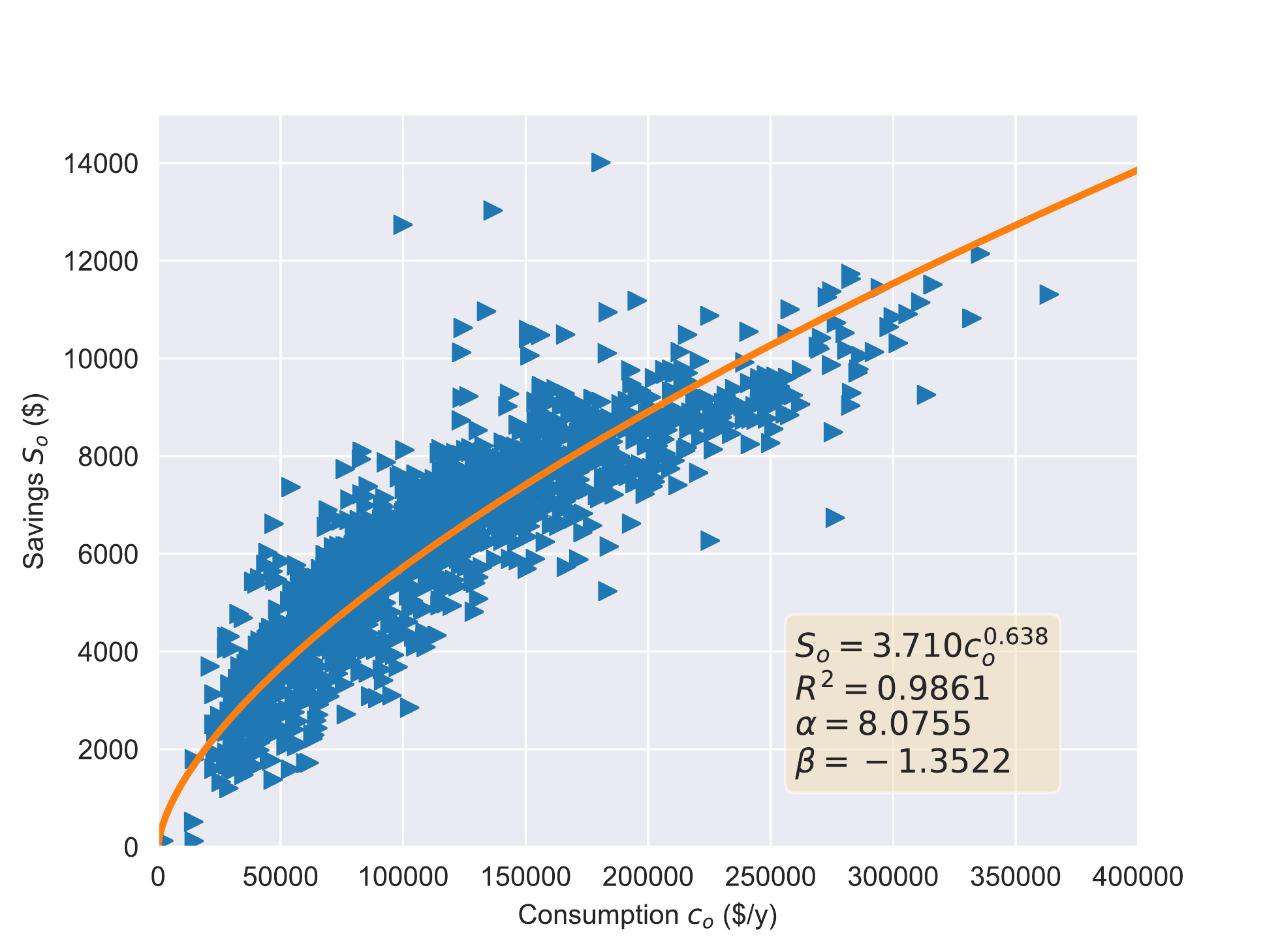}
    \caption{Utility of per capita savings exponential calibration}
    \label{fig:Fig1_SavingsCalibration}
\end{figure}

The exponential calibration of the utility of precautionary savings is shown in Figure \ref{fig:Fig1_SavingsCalibration} with coefficients $a=3.710, b=0.638$, which a coefficient of determination of $R^2=0.9861$. 

\section{California Labor Statistics}
\label{sec:CaliforniaLabor}
\begin{figure}[h!]
    \centering
    \includegraphics[width=\linewidth]{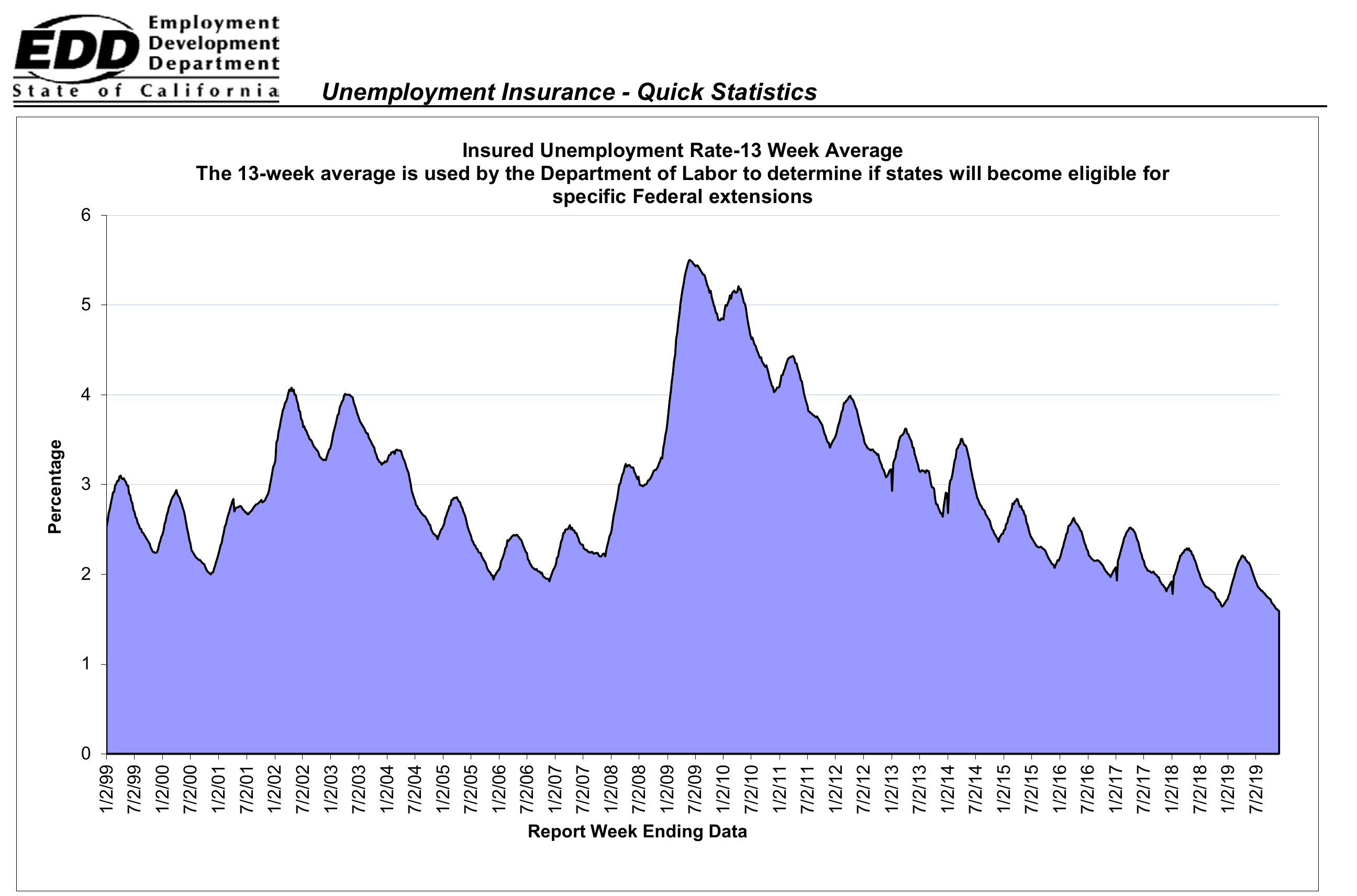}
    \caption{Insured unemployment rate (IUR) using a 13-week average from 1999 to 2019 for California according to the Employment Development Department (EDD), State of California, \url{https://www.edd.ca.gov/about_edd/quick_statistics.htm\#UIStatistics} }
    \label{fig:UIR_California}
\end{figure}

\begin{figure}[h!]
    \centering
    \includegraphics[width=\linewidth]{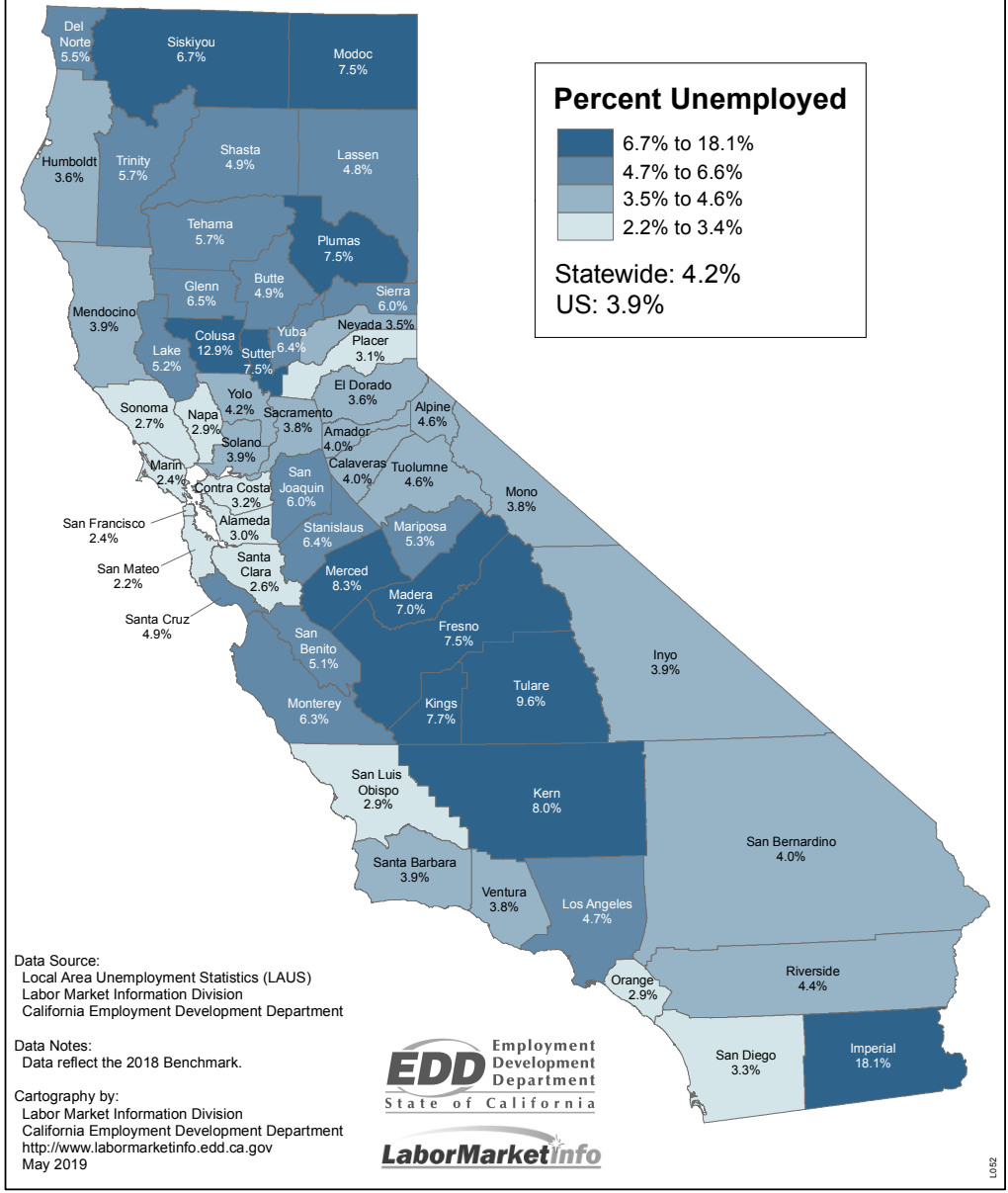}
    \caption{Map of annual average unemployment rate in California by county in 2018 according to the Employment Development Department (EDD), State of California, \\ \url{https://www.labormarketinfo.edd.ca.gov/file/Maps/County_UR_2018BM2018.pdf}}
    \label{fig:CountyUnemploymentRate}
\end{figure}
\section{California Poverty Rate}

\begin{figure}[h!]
    \centering
    \includegraphics[width=0.9\linewidth]{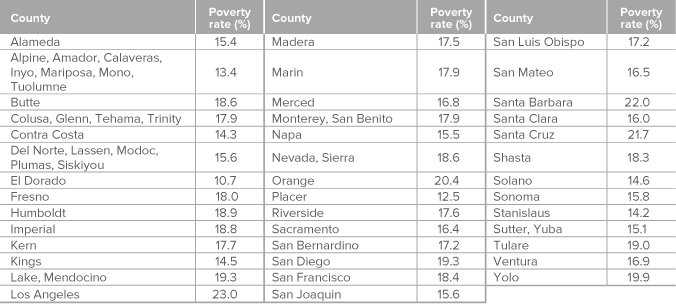}
    \caption{Poverty rates for California counties according to the California Poverty Measure (CPM) courtesy of the Public Policy Institute of California (PPIC) and Stanford's Center on Poverty and Inequality, \url{https://www.ppic.org/publication/poverty-in-california/}}
    \label{fig:PovertyCalifornia}
\end{figure}

\end{document}